# Polarization-resolved haze in transparent wood – towards wooden glass

Zheheng Song[1*], Hui Chen[2], Mathias Nero[3,4], Jingjian Zhou[1], Ravi Shanker[1], Tom Willhammar[3], Lars Berglund[2] and Ilya Sychugov[1*]

[1] Department of Applied Physics, KTH Royal Institute of Technology, Stockholm 11419, Sweden

[2] Department of Fibre and Polymer Technology, Wallenberg Wood Science Center, KTH Royal Institute of Technology, Stockholm 10044, Sweden

[3] Department of Chemistry, Stockholm University, Stockholm 10691, Sweden

[4] Department of Fibre and Polymer Technology, Fibre Processes, KTH Royal Institute of Technology, Stockholm 10044, Sweden

*ilyas@kth.se *zheheng@kth.se

## Abstract

Transparent wood (TW) composites possess residual haze that limits image clarity despite their high light transmission, yet the physical origin of this scattering has not been established. Here, polarization-resolved measurements reveal a pronounced polarization dependence of scattering by a wood–polymer composite with improved refractive index matching. Polarization- and spectrally-resolved haze as low as 21% (40-45 % at 550 nm for unpolarized light) with transmittance of ~92% for polarization perpendicular to the fibers was detected. It is found that the low-haze spectrum closely follows an inverse-square relation over 400–800 nm range, well described by a weak-index-contrast Rayleigh–Gans–Debye (RGD) model invoking subwavelength birefringent heterogeneities within the cell wall. This physical mechanism was previously obscured by haze-limiting strong scattering from debonding gaps or refraction at the lumen-cell wall interface. Scanning electron diffraction independently reveals spatially distinct domains with markedly different cellulose diffraction responses within the infiltrated cell walls. The observed polarization-dependent refractive-index mismatch is attributed to inherent cellulose birefringence. Time-resolved measurements further support nearly scattering-free quasi-ballistic transport of the sampled transmitted light. It is demonstrated that selecting only a preferential polarization markedly improves image visibility through TW without further material modification. These findings connect polarization-dependent haze to the internal structure of transparent wood and identify ways to engineer a refractive-index match in hierarchical anisotropic composites towards glass-like clarity.

## Introduction

Since early use of transparent wood (TW) for anatomical observation in 1992 by removal of light-absorbing components (lignin) and infiltration with a refractive index-matching polymer[1], this concept has developed into a structural wood–polymer composite[2,3]. The combination of mechanical and optical properties makes this renewable resource potentially attractive for optical windows, optoelectronic substrates and building-integrated photovoltaics[2–5]. As monolithic glass is a material with high embodied energy and relies on finite raw-material resources (riverbed sand), developing engineered composites is an important research direction. Both the amount (transmittance) and path (scattering/haze) of transmitted light through TW matter. As summarized in recent reviews, values for both quantities have typically been high[6–11]. While high transmittance is desirable for the "wooden glass" concept, substantial haze redistributes transmitted light away from the ballistic direction, thereby degrading image clarity[12].

The crystalline cellulose framework serves as a mechanically reinforcing skeleton for this composite. Therefore, when considering achieved low haze values for transparent wood (TW), it is necessary to limit examples to the ones with a substantial amount of cellulose in the mixture. Thus, extensive delignification leading to nearly complete removal of cell wall material, essentially reducing the original framework to pulp, can hardly be considered representative[13,14], as pointed out in a review[8]. Similarly, transversely cut wood samples with light propagation along the fiber direction may yield low haze values, yet such a composite is void of any mechanical strength advantages of wood[15]. For the same reason, very thin samples of only a few hundred microns thick cannot be representative, as haze increases with thickness[16–18]. On the measurement side, correct evaluation of haze values requires a modified protocol (ASTM D1003, originally developed for low scattering plastics)[19], otherwise a systematic error towards lower haze may appear due to small integrating sphere size and uncontrolled aperture size[20,21]. With all the filters applied to hundreds of papers on TW from the reviews above, we can state that with rare exceptions measured under above conditions[20,21], the lowest haze values demonstrated so far hover around 50% for samples at least 1 mm thick with a meaningful amount of cellulose. Clearly, this value needs to be reduced at least an order of magnitude for any practical advances of the "wooden glass" concept.

Limited progress in this direction partly reflects the complexity of light scattering in TW composites, which remains not fully understood. Indeed, this material contains potential scattering sources across multiple length scales. Interfaces around vessels and cell lumens, residual air gaps and heterogeneities within the cell wall can all redirect light through local refractive-index differences[22–24]. Their overlapping contributions make haze difficult to unequivocally assign to a particular structure or size and, instead, may require the sequential unraveling and mitigation of individual sources. One of the first major contributions recognized was air gaps between the lumen-infiltrated polymer and the wood cell wall. Those form during polymerization shrinkage and can be avoided by chemical bonding between the filler and the host cell wall[20–22]. Next, cell-wall swelling was used to promote polymer infiltration into the wall, making the resulting composite more optically homogeneous (largely preserving strength, increasing toughness at the cost of roughly halving stiffness)[25].

On the characterization side, photon-budget and angle-resolved measurements revealed directional scattering[12,23]. Earlier polarization-resolved studies demonstrated that scattering and depolarization depend on the incident polarization relative to the wood fibers[13,26]. However, the high overall haze of those composites, together with overlapping scattering contributions and depolarization, made it difficult to isolate and quantitatively relate their haze spectra to specific scattering structures[12,26]. Time-resolved measurements have further shown predominantly forward, quasi-ballistic transport in TW composites, in contrast to diffusive propagation in native wood[27]. All these advances pointed to light refraction at the cell wall/polymer interface as a major remaining source of scattering.

Recently, fine-tuning of the filler refractive index through poly(methyl methacrylate) (PMMA) doping with phenanthrene was realized as a model system to achieve a better refractive index matching with the cell wall[28]. As a result, this well-developed polymer infiltration method could be applied as other important parameters, such as viscosity, were not much affected. By suppressing refraction in this way, the resulting haze (at 550 nm) was pushed down to 40-50%. In the red spectral range, the value dropped below ~40% for a 1-mm-thick sample, which is one of the lowest values ever measured with a verified technique on cellulose-rich and relatively thick TW samples. The improved low-haze material creates an opportunity to resolve the optical anisotropy that persists after interface refinement and homogenization. Because the refractive index of cellulose depends on polarization[24,29,30], orthogonal incident polarizations probe different optical contrasts within the same composite. Resolving these responses can expose scattering channels hidden by polarization averaging and identify how its spectral dependence relates to internal structure. Connecting a verified polarization-dependent haze spectrum with a physical scattering model and independent structural evidence can therefore guide both the interpretation of residual scattering and the design of a clearer transparent wood.

Here, we investigated residual haze through polarization-resolved spectroscopy following updated haze measurement protocol for TW[19]. Complementary measurements with a narrow 0.5 mm beam and a calibrated haze reference were used to verify the polarization response. It was found that for a specific polarization in the red spectral range TW haze can be as low as ~20%, setting a new milestone in the TW material development. Time-resolved transmission measurement probed photon transport and confirmed substantially reduced scattering in this optical domain. The observed spectral dependence was analyzed within a weak refractive index-contrast Rayleigh-Gans-Debye (RGD) model, linking the measured haze to a characteristic scattering feature size[31–33]. Scanning electron diffraction (SED)[34] resolved local structural heterogeneity within the infiltrated cell walls, consistent with the ~50–200 nm scale of refractive-index heterogeneities deduced from the optical measurements. Together, these measurements provide a consistent physical basis for improving image clarity through polarization selection and refractive-index engineering. The pronounced polarization dependence shows that residual haze depends not only on average refractive-index matching but also on the anisotropic optical response of the cellulose framework, whose effective birefringence is modulated by microfibril orientation. This identifies cellulose anisotropy and fibrillar organization as key parameters for the optical design of low-haze anisotropic composites.

## Results

### Optical transparency of the modified wood composite

The newly developed transparent wood (TW), with a thickness of 1 mm, retains remarkably clear visibility at a large object-to-sample separation[28]. A building approximately 200 m behind the specimen remains readily recognizable, with its curved façade and horizontal architectural features clearly discernible through the composite (Fig. 1a). The material also allows clear observation of a color target placed in contact with the specimen (Fig. 1b, inset). All nine colored circles are readily distinguished, with distinct colors and well-defined outlines. The reduced haze enables the modified TW to preserve image information over both near-contact and long-distance viewing conditions (with some visible distortions for the latter).

The composite was prepared using the multistep optical-design framework previously reported[28], with alkaline swelling and acetylation of the wood template followed by infiltration with a refractive-index-adjusted polymer. The fabrication sequence is shown in Fig. S1, while Fig. 1d illustrates the hierarchical structure of the final composite. Polymer fills the large anatomical cavities and penetrates the cellulose-rich walls, forming a hierarchical composite with structural features spanning multiple length scales. Annular dark-field scanning transmission electron microscopy (ADF-STEM) and local nanobeam electron diffraction on thin microtomed lamellas reveal distinct microscopic environments within the transparent composite (Fig. 1c). Polymer-rich regions (orange box) exhibit predominantly diffuse electron scattering around the transmitted beam, whereas a representative region in the secondary cell wall (green box) shows pronounced diffraction reflections, consistent with a crystalline cellulose-rich environment. Spatial variations in cellulose diffraction within the cell wall are examined separately in the SED analysis below. The contrasting diffraction responses further indicate substantial local structural heterogeneity within the infiltrated composite.

The unpolarized measurements establish the photon budget of the composite (Fig. 1b). At 550 nm, total transmittance reaches approximately 90%, while reflectance remains below 8% and absorptance is only a few percent. Towards 800 nm, transmittance approaches 92% and absorptance becomes negligible. In contrast to the relatively weak spectral variation in total transmission, haze shows a pronounced wavelength dependence, decreasing from approximately 60% near 400 nm to 43% at 550 nm and reaching approximately 30% near 800 nm. At 550 nm, the haze accounts for approximately 38% of the incident light, compared with approximately 51% in the remaining transmitted component. The photon budget therefore shows that high transmission and low absorption coexist with substantial wavelength-dependent haze. Polarization-resolved measurements were then used to determine how this residual haze depends on the orientation of the incident electric field.

**Resolving polarization-dependent haze**

The reduced haze of the present composite makes this anisotropic response more clearly accessible. Spectral measurements following the integrating-sphere protocol developed for transparent wood[19] resolve a systematic polarization dependence over 400–800 nm (Fig. 2a). The angle $\theta$ denotes the orientation of the incident electric field relative to the fibers, with 0° and 90° corresponding to parallel and perpendicular polarization, respectively (inset). The haze spectra separate progressively as the incident polarization rotates from parallel to perpendicular to the fibers. Haze decreases with increasing wavelength for all measured angles, with a substantially stronger decrease for perpendicular polarization. Near 800 nm, haze reaches approximately 21% for perpendicular polarization, approaching the level of soiled glass[35]. Spectra at the other polarization angles follow the same systematic ordering. In contrast, total transmittance depends only weakly on polarization at a given wavelength. The pronounced haze contrast therefore primarily reflects polarization-dependent angular redistribution of transmitted light rather than a substantial change in the total transmitted fraction. We further examined the polarization response using an independent 532 nm laser setup with a nominal 500 μm incident beam and a corresponding integrating sphere exit aperture (Fig. S2a). The narrower beam reduces the influence of a spread beam width and enables local measurements and systematic spatial scans. A nominal ASTM haze standard with a certified haze value of 26.99% gives approximately 27.1% and 27.3% for the two orthogonal polarizations and remains nearly constant across the sampled positions (Fig. S2b, c), confirming that the setup itself does not introduce any polarization-dependent variation.

Using the same optical setup, the TW angular scan in Fig. 2b reproduces the polarization ordering observed in the spectral measurements. Haze decreases from approximately 55.9% at 0° to 40.7% at 90°, with intermediate angles following the same overall trend despite small local fluctuations. Total transmittance remains within approximately 86–90% and shows no comparable angular dependence. Circular polarization provides an additional intermediate response in the same configuration: the measured haze is approximately 47.5%, while the unpolarized reference is approximately 45%, both lying between the parallel- and perpendicular-polarization responses. These measurements independently confirm a pronounced polarization dependence of haze, whereas total transmission remains nearly polarization independent. The TW spatial scan maintains the same relative haze levels at all sampled positions (Fig. S2d). The parallel-polarization haze lies approximately within 52–56%, while the perpendicular response remains around 40–43%. The 45° and circular-polarization measurements occupy an intermediate range around 46–48%. Local variation is evident, but it remains much smaller than the polarization-induced separation. Thus, the polarization-dependent haze is a spatially robust feature of the composite rather than a local effect.

A comparison TW specimen without refractive-index modification of the poly(methyl methacrylate) (PMMA) matrix retains high transmittance but remains strongly hazy across the measured spectral range (Fig. S3). The angular scan shows that haze decreases from approximately 86% for parallel polarization to about 79% for perpendicular polarization, while total transmittance remains around 88–90%. The spectral measurements similarly show haze broadly in the 80–90% range for all polarization angles, with only a modest separation between them. Thus, some polarization dependence persists even in the refractive-index-mismatched composite, but both polarization states remain strongly hazy. Refractive-index fine-tuning is therefore essential for suppressing the overall optical mismatch and enabling the low-haze perpendicular-polarization channel observed in the optimized composite.

**Temporal profiles of transmitted light**

Time-resolved measurements were used to examine whether the strong angular redistribution observed in the haze measurements is accompanied by temporal broadening. The sample was exposed to 790 nm light of polarization perpendicular to fibers, while the collection aperture (slit) was translated laterally by $\Delta x$ to sample transmitted photons away from the central propagation direction (Figs. 3 and S4a). As the collection aperture moves away from the center, the detected intensity decreases strongly, reaching approximately 19% of the central intensity at $\Delta x$ = 0.70 mm (Fig. 3a, c). Despite this substantial loss in collected intensity, the corresponding temporal profiles remain compact and symmetrical, with no systematic broadening as the displacement increases (Fig. 3b). Mean temporal FWHM values remain approximately 5–8 ps across the entire scan, with no systematic increase as $\Delta x$ increases (Fig. 3d). This compact temporal response contrasts with the pronounced temporal broadening previously reported for native wood[27].

The absence of increasing temporal broadening despite the strong spatial decrease in intensity indicates that the distribution of arrival times does not become substantially broader at the sampled lateral positions. This behavior is consistent with predominantly ballistic or quasi-ballistic transport in the collected light. A separate comparison further shows that refractive-index matching influences the temporal response (Fig. S4b,c). Pure PMMA exhibits the narrowest temporal profile, TW with the conventional PMMA matrix ($n = \sim 1.49$) the broadest, and refractive-index-modified TW ($n = \sim 1.53$) an intermediate width. The narrower profile is consistent with reduced scattering-induced temporal broadening after refractive-index modification.

**Spectral model and nanoscale structural observations**

The perpendicular-polarization haze exhibits a close inverse-square spectral dependence over 400–800 nm (Fig. 4a). The data are well described by an offset relation of the form $H(\lambda) = H_0 + C\lambda^{-2}$, capturing the pronounced wavelength dependence with a compact empirical description ($H_0 \approx 0.08$, which can originate from e.g., surface roughness). This inverse-square contribution is consistent with the weak-scattering RGD parameterization adopted here for finite-sized cylindrical refractive-index heterogeneities[32,33]. The parallel-polarization spectrum is almost spectrally independent and cannot be fitted with $\lambda^{-2}$ relation. The polarization dependence is also evident in the far-field scattering patterns (Fig. S5). For vertical fibers, parallel polarization produces a horizontally elongated intensity distribution, whereas perpendicular polarization retains a more concentrated central feature. Rotating the fibers rotates the dominant elongation accordingly. The preferred spreading direction therefore follows the material orientation, confirming that the scattering response is strongly anisotropic.

Figs. 4b, c illustrate the refractive-index picture used to interpret this anisotropy. Light propagates across the aligned structure, while the electric field probes either the response parallel or perpendicular to the fibers. Representative refractive indices are ~1.53 for the infiltrating polymer, ~1.52 for the perpendicular cellulose-framework response, and ~1.58 for the parallel response, corresponding to mismatch magnitudes of approximately 0.01 and 0.05, respectively. The birefringence values adopted here are representative of wood, whereas the intrinsic birefringence of aligned cellulose has been estimated even higher at $\Delta n = 0.09$[29,30]. Indeed, naturally well-aligned ramie

fibers exhibit birefringence $\Delta n = 0.07 - 0.08$ in the visible and NIR ranges[36]. In general, the effective birefringence of the cell wall depends on the microfibril angle relative to the wood fiber axis and on the orientations and relative thickness of the cell-wall layers[30]. Consequently, variation in microfibril orientation can modify the refractive-index mismatch experienced by the two incident polarizations. Given the dependence of effective birefringence on microfibril orientation, the values adopted here are literature-based estimates representative of comparable wood systems. For example, refractive index measured along the fiber direction (both transverse components are perpendicular to the fiber axis) gives ordinary index 1.525 ± 0.008 for delignified balsa, supporting the selected value of ~1.52 for this polarization (across the fibers)[24]. As the diffraction data shown below indicate nanofibrillar rearrangement and a narrower orientational distribution after swelling, these data for non-acetylated, non-swollen TW may not be directly applicable to estimating the extraordinary refractive index along the fibers. Instead, typical birefringence value for wood $\Delta n = 0.06$ was applied.

Within the adopted cylinder model, the fitted inverse-square coefficient links the refractive-index contrast to a characteristic scattering scale perpendicular to the fibers as the product $d(\Delta n_{\perp})^2$ can be extracted. Using a 1 mm model path length and an assumed scattering-region volume fraction of 1/8 (lumen diameter of ~30 µm, cell wall thickness of ~5 µm, and equal cellulose/polymer fractions in the swollen cell wall), mismatch magnitudes $\Delta n_{\perp}$ of 0.01 and 0.02 correspond to effective scatterer diameters from 200 to 50 nm, respectively. These values represent conditional estimates of the characteristic structural scale rather than a directly measured size distribution. The full derivation and parameter dependence are given in Supplementary Note 1.

**Nanoscale structural heterogeneity revealed by SED**

To examine the nanoscale structural heterogeneity associated with this scattering picture, SED was used to compare TW prepared with and without NaOH treatment. SED allows the creation of virtual maps that reveal the crystalline texture at nanoscale resolution. The continuous and binary cellulose diffraction maps reveal pronounced spatial heterogeneity within the treated wall (Fig. 4d). Brighter regions in the continuous map correspond to more pronounced cellulose diffraction, whereas the binary map distinguishes domains with and without any detectable diffraction signal from cellulose[25]. The SED results from NaOH-treated material show irregular patches of contrasting signal, indicating substantial local variation in the crystalline texture within the infiltrated cell wall. Virtual bright-field images locate the regions used for the diffraction comparison (Fig. S6). Reference TW, prepared without NaOH treatment, shows comparatively diffuse arc-like cellulose reflections across the selected region, whereas the treated sample contains sharply defined reflections adjacent to positions with weak or undetectable diffraction.

Representative diffraction patterns are shown in Fig. S7a, b. The azimuthal FWHM distributions of both the cellulose $(200)$ and $(110)/(1\bar{1}0)$ reflections shift towards narrower reflections after NaOH treatment (Fig. S7c). At the same time, the fraction of sampled positions with detectable reflections decreases from 98% in the reference TW to 51% in the treated sample, while simultaneous detection of both reflection families decreases from 90% to 9% (Fig. S7d). The SED data show that the treated cell wall contains fewer locations with detectable diffraction, but sharper diffraction signals where reflections are observed. The larger area and local binary maps locate the heterogeneous cellulose diffraction response within the cell wall (Fig. S8). Detectable diffraction is comparatively continuous in the reference TW and more spatially interrupted in the NaOH-treated TW. These maps establish spatial variation in crystalline texture, providing structural evidence consistent with the proposed wall-internal scattering picture. The acquisition and analysis procedures are described in the Supplementary Methods.

**Improved image clarity through polarization selection**

The polarization-dependent haze directly affects image formation through the composite (Fig. 5). A logo and color target were viewed at an object-to-TW separation of approximately 30 cm in room light. Through TW alone, the logo remains recognizable, although its lettering and fine features are blurred by residual haze (Fig. 5a). With the linear-polarizer transmission axis parallel to the fibers, the image becomes more strongly blurred (Fig. 5b), whereas perpendicular polarization produces clearer lettering and sharper boundaries between the colored regions (Fig. 5c). An additional comparison between close contact and an approximately 30 cm object separation further illustrates this effect (Fig. S9). Image detail is readily preserved in contact, whereas increasing the separation makes the polarization-dependent loss of clarity more apparent. Perpendicular polarization consistently improves feature visibility, demonstrating that the low-haze polarization channel can be directly exploited to improve image clarity without further modification of the composite.

## Discussion

The birefringent crystalline cellulose-rich domain framework is a physical basis for the polarization-dependent haze. Because the polymer refractive index in our case is closer to the cellulose response for polarization perpendicular to the fibers, the refractive-index mismatch is smaller in this orientation and larger for polarization parallel to the fibers. The comparison with conventional PMMA further shows that refractive-index modification is essential: although transmittance remains high without index tuning, both polarization states retain substantial haze. Refractive-index engineering therefore suppresses the overall optical mismatch while preserving the intrinsic birefringence of cellulose, enabling a distinct low-haze polarization channel. These results indicate that optical design should address polarization-dependent refractive-index mismatch rather than optimize only polarization-averaged transmission for TW.

The spectral dependence provides a key constraint on the scattering regime. For perpendicular polarization, the haze follows an offset inverse-square $\lambda^{-2}$ relation over 400–800 nm, distinctly different from the $\lambda^{-4}$ dependence expected in the classical Rayleigh limit. The inverse-square dependence is instead consistent with weak-contrast scattering from finite-sized refractive-index heterogeneities. The RGD model is used here as the weak-contrast approximation to the more general particle-scattering problem. A standard Mie calculation would require an assumed spherical geometry and size distribution that are not independently supported for the wall-internal heterogeneities. By contrast, the RGD framework for cylinders is compatible with the estimated small index contrast for perpendicular polarization and retains an explicit dependence on scatterer geometry (cylinder diameter). The fitted spectrum can therefore constrain a characteristic scattering scale. Should air-filled thin voids parallel to the fiber exist, the polarization-resolved response would be the opposite. With a large refractive index contrast (1 vs. 1.5) the perpendicular polarization would induce stronger polarization across such gaps, resulting in a more pronounced scattering. The parallel-polarization spectrum is not interpreted using this model because it is almost wavelength-independent and, thus, scattering in that case is mainly refraction-limited.

The RGD model places the characteristic scattering scale well below the dimensions of fiber lumina and vessels, pointing instead to heterogeneities within the fiber wall. These heterogeneities may include polymer-rich or mixed cellulose–polymer regions whose refractive-index response differs from that of the surrounding birefringent framework. For the adopted range of refractive-index contrast, scattering fraction and optical path length, the fitted spectrum corresponds to an effective structural scale of approximately 50–200 nm. This result links the macroscopic haze spectrum to nanoscale structural heterogeneity within the infiltrated wall. Electron microscopy shows continuous contact between the infiltrated polymer and the cell wall, with no obvious interfacial gaps in the examined regions (Fig. 1c). These observations independently argue against widespread interfacial voids as the dominant source of residual haze. SED further reveals spatial heterogeneity in the cellulose diffraction response within the infiltrated cell wall (Figs. 4d and S8), supporting the proposed wall-internal scattering mechanism.

The SED comparison links the proposed scattering domains to structural changes induced by NaOH treatment. Alkaline swelling increases the accessibility and nanoscale porosity of the cell wall, allowing the infiltrating polymer to penetrate regions that would otherwise remain poorly filled[25,28]. This improved infiltration reduces strong refractive-index contrasts associated with residual air-filled nanovoids and poorly matched interfaces and is therefore essential for achieving the low overall haze of the composite. SED nevertheless shows that the infiltrated cell wall retains pronounced nanoscale heterogeneity in its cellulose diffraction response. The response combines fewer detectable diffraction sites with sharper reflections where diffraction remains observable, consistent with increased orientational and rotational order among the diffracting cellulose nanofibrils. Low-signal regions may correspond to polymer-rich or mixed cellulose–polymer environments. NaOH-assisted infiltration therefore suppresses the much stronger scattering associated with unfilled pores and interfaces, while leaving smaller local refractive-index variations within the infiltrated wall that can contribute to the residual haze. Cellulose birefringence reveals the remaining local refractive-index variations as a polarization-dependent scattering response. The design objective is therefore not to eliminate the swelling treatment, but to preserve its infiltration benefit while further reducing refractive-index variations between neighboring wall domains. This shifts the optimization from improving infiltration alone towards engineering a more optically homogeneous cell-wall environment.

The temporal profiles remain symmetrical (Fig. 3b) and show no systematic broadening with increasing lateral collection displacement, despite a substantial decrease in detected intensity. This is in contrast with most previous time-resolved measurements on TW, as those typically featured a prolonged tail from the scattering component[23,27]. The composite therefore retains a compact temporal response under favorable polarization, indicating the lack of zigzag, diffusive photon trajectories caused by multiple scattering.  Indeed, for ~20% haze one can estimate the probability of two scattering events as negligible (few percent) in such 1 mm thick samples for transmitted photons.

The revealed mechanism suggests a clear materials-design strategy towards “wooden glass”. Further reduction of residual scattering may be achieved by: (i) using a birefringent polymer matrix whose refractive indices are better matched to the cellulose framework in both polarization directions, for example by aligning nanorods or liquid-crystalline components before polymerization[37]; (ii) reducing the size of the refractive-index heterogeneities through further control of cell-wall swelling and polymer infiltration; (iii) selecting wood structures with less circular lumen geometries, which would reduce refraction at lumen-wall interfaces for a given refractive-index mismatch[38]; and (iv) using wood species with broadly distributed microfibril angles at nanoscale (compression or juvenile wood), yielding a nearly polarization-independent optical response. At the device level, polarization selection could be combined with polarization-selective optoelectronic elements to utilize the stronger scattering component rather than simply rejecting it. Organic photovoltaic cells with controlled polarization sensitivity provide a precedent for such selective conversion[39], while polarization-sensitive organic photodetectors demonstrate a related strategy for optical sensing[40]. These approaches could enable transparent wood systems that combine improved optical clarity with polarization-selective energy harvesting.

Overall, polarization- and wavelength-resolved measurements revealed how residual haze in transparent wood emerges from anisotropic refractive-index mismatch within its hierarchical internal structure. The strong polarization contrast, inverse-square spectral dependence and nanoscale structural heterogeneity together connect the macroscopic optical response to wall-internal refractive-index variations. These findings establish polarization-resolved scattering as a powerful approach for probing and engineering optical mismatch in hierarchical anisotropic composites. Polarization selection already provides an immediate improvement in image clarity, while further control of wall-internal refractive-index heterogeneity offers a path towards still lower haze and brings transparent wood closer to the optical clarity envisioned for "wooden glass".

## Acknowledgements

The authors are grateful for funding from Knut and Alice Wallenberg foundation (Dnr 2021.0311). Z.S. acknowledges funding support from the China Scholarship Council (CSC).

## Materials and methods

### Materials

TW was prepared from balsa veneers with a nominal starting thickness of 1 mm following the multistep protocol developed by Chen et al.[28]. The veneers were delignified, swollen in aqueous NaOH and acetylated before solvent exchange and vacuum infiltration with prepolymerized methyl methacrylate (pre-MMA). Phenanthrene was dissolved in the pre-MMA to adjust the refractive index of the resulting PMMA matrix. The infiltrated templates were cured between glass slides, resulting in samples of thickness ~1 mm with wood volume fraction of 8.0 ± 0.7%. Detailed preparation conditions are provided in the Supplementary Methods.

### Methods

Spectral transmittance and reflectance were measured using a laser-driven xenon plasma light source (Energetiq EQ-99), a monochromator (SP2150i, Princeton Instruments) to select an incoming wavelength, a 6-inch integrating sphere (Labsphere) connected to a Peltier-cooled CCD via a spectrometer (Princeton Instruments). Monochromatic wavelength-by-wavelength excitation was used below 500 nm, and a broadband light was applied above 500 nm (stitching can be noticed in Fig. 4a). Polarization-resolved spectra were acquired over 400–800 nm range, with a linear polarizer controlling the electric-field angle relative to the fibers. A quarter-wave plate provided circular polarization, and the setup response was checked using a nominal 26.99% haze standard at different polarization angles and positions. Transmittance was referenced to the incident light in each selected polarization state. A separate well-collimated beam from a 532 nm laser was directed to an aperture with a diameter of 500 μm to narrow the incident beam size. In this case a custom 7.5 mm diameter exit aperture was used in the integrating sphere for polarization-angle and spatially resolved haze measurements. The perpendicular-polarization haze spectrum was fitted with an offset inverse-square relation and interpreted using a weak-contrast RGD model; details are provided in Supplementary Note 1. Time-resolved transmission was measured using a Ti:sapphire laser (Mira Optima 900-F, Coherent) delivering approximately 150 fs long pulses at 790 nm central wavelength and 76 MHz repetition frequency, together with a Hamamatsu C5680 streak camera equipped with an M5675 synchroscan unit. Measurements were performed with a 200 ps sweep range. A micrometer stage translated the collection slit laterally while the illuminated sample position remained fixed. Temporal profiles were normalized and aligned at their peaks for shape comparison, and the reported temporal widths include the instrument response (~4 ps), unless specified otherwise when deconvoluted. The laser polarization was perpendicular to the fiber direction. For electron diffraction, transparent wood samples were sectioned transversely into a thickness of around 200 nm using an ultramicrotome (Leica Ultracut UCT) at room temperature. Data were acquired using a Thermo Fisher Themis-Z transmission electron microscope operating at 300 kV with a CheeTah M3 hybrid electron camera (512 x 512 pixels) from Amsterdam Scientific Instruments (beam current 2 pA; convergence angle 0.09 mrad; dwell time 5 ms; step size 30 nm). SED data analysis of the cellulose $(200)$ and $(110)/(1\bar{1}0)$ reflections was used to compare diffraction peak widths, detection frequencies, and spatial heterogeneity using a custom-designed post-processing routine, see Supporting Information for details.

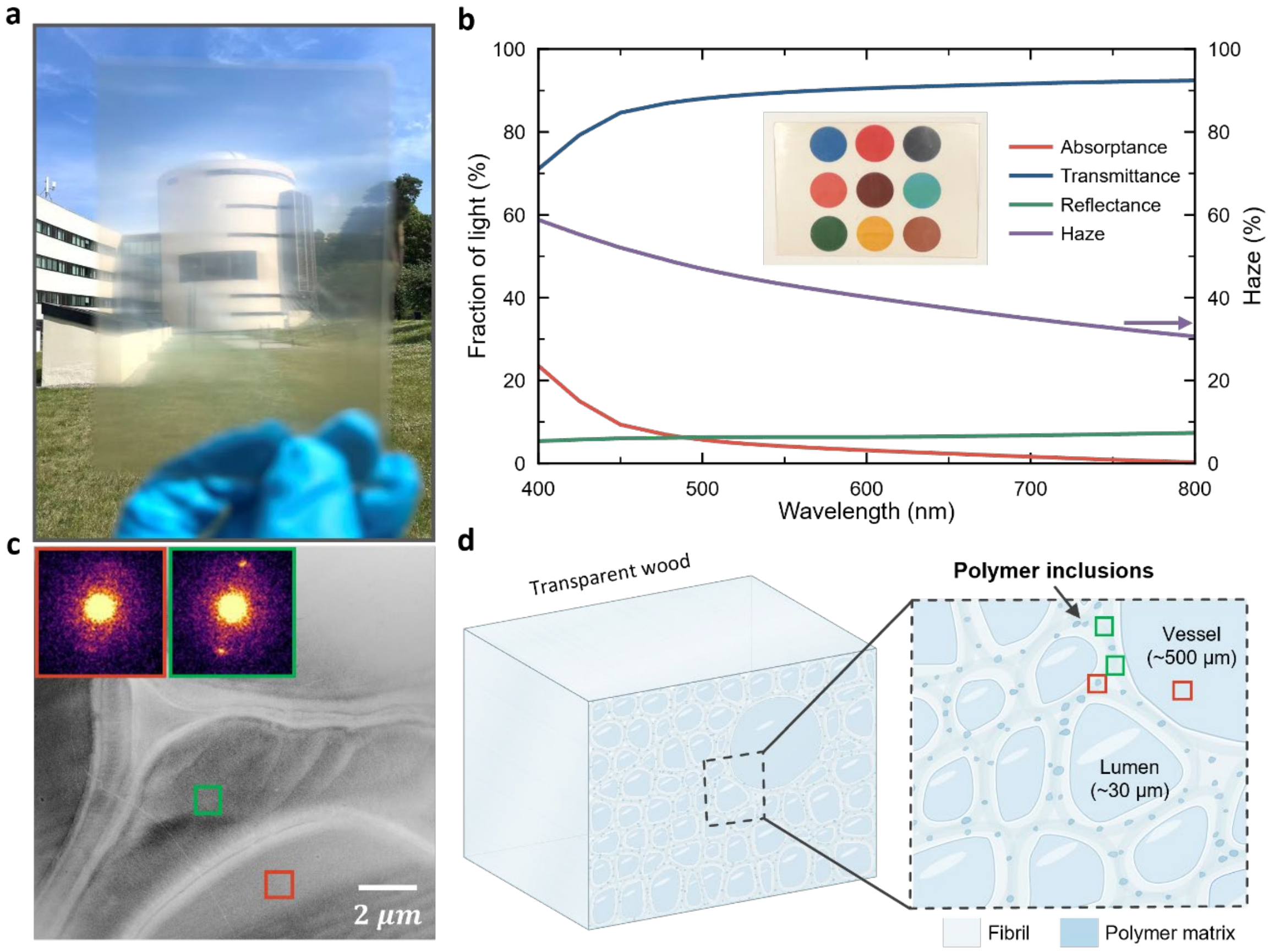

**Figure 1. Optical transparency and local structural heterogeneity of low-haze transparent wood.** **a**, Photograph through 1-mm-thick transparent wood (TW), showing a building approximately 200 m away. The composite comprises a 1 mm balsa wood template (wood volume fraction, ~8.0%) and a phenanthrene-modified PMMA matrix. **b**, Spectral photon budget under unpolarized illumination. Inset, image was viewed through TW in contact with the specimen. **c**, Annular dark-field scanning transmission electron microscopy (ADF-STEM) image showing selected regions in the lumen (orange box) and in the cell wall (green box). Inset, representative electron diffraction patterns from the regions marked in orange and green, respectively, with matching frame colors. Scale bar, 2 μm. **d**, Schematic of the hierarchical structure of TW, illustrating polymer-filled vessels (~500 μm) and lumen (~30 μm), together with polymer inclusions within the cell walls. Schematic not to scale.

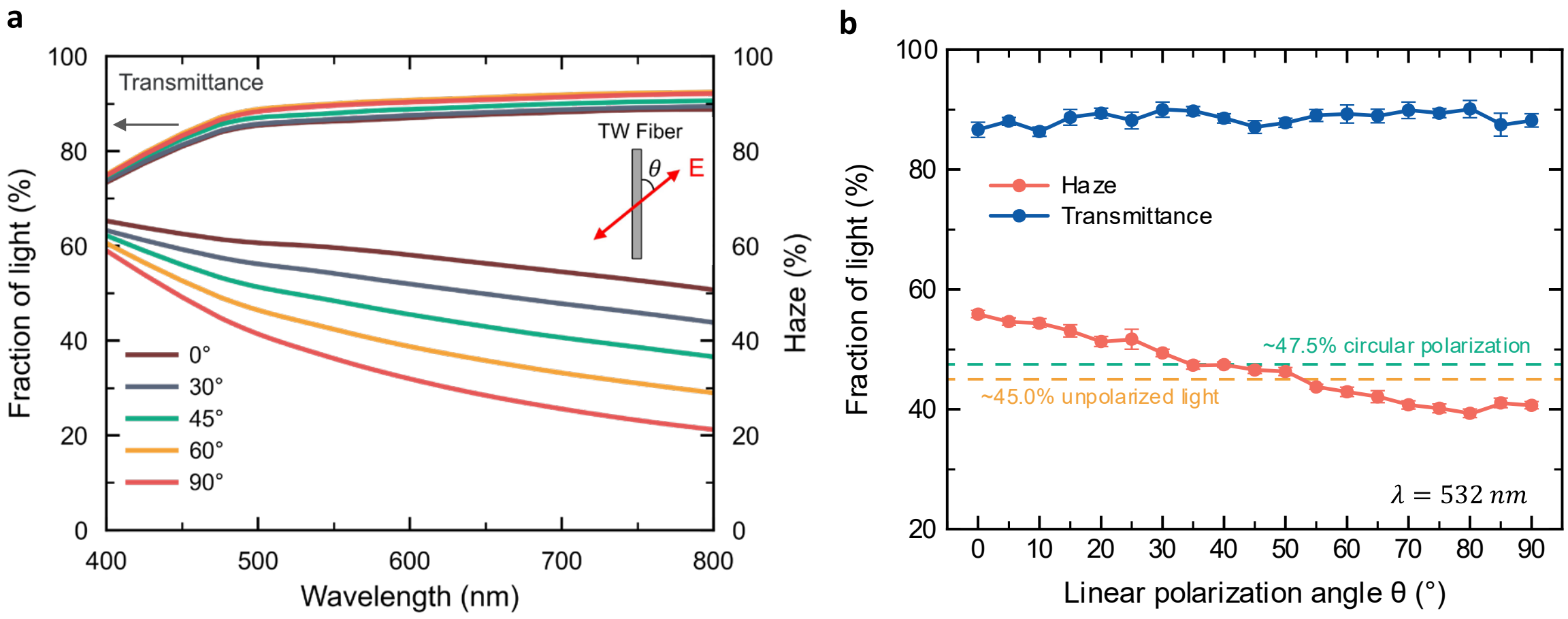

**Figure 2. Polarization-resolved transmittance and haze of transparent wood. a,** Transmittance (left axis) and haze (right axis) spectra at selected polarization angles. The inset defines $\theta$ between the incident electric field E and the wood fibers. Perpendicular polarization ($\theta = 90°$) yields haze as low as approximately 21% near 800 nm. **b,** Angular dependence of transmittance and haze measured at 532 nm using an independent 500 μm beam setup. Dashed lines indicate circularly polarized and unpolarized haze values.

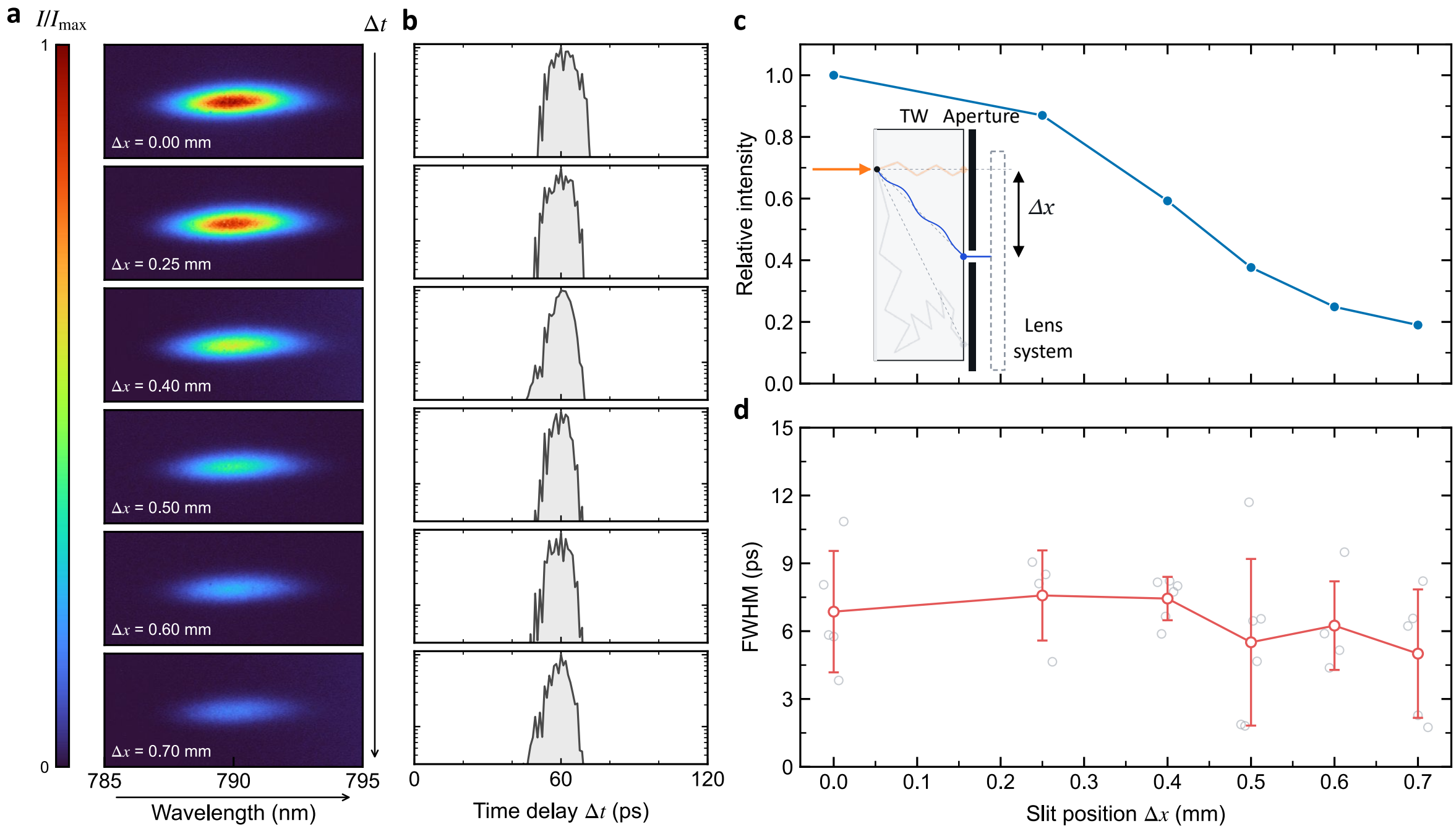

**Figure 3. Time-resolved characterization of photon transport through transparent wood. a,** Streak-camera images at increasing slit (aperture) displacements $\Delta x$ under excitation at ~790 nm with polarization perpendicular to the fibers. **b,** Corresponding normalized temporal profiles (log-scale), aligned at their peaks. **c,** Collected intensity normalized to that at $\Delta x = 0$ versus slit displacement. The inset illustrates the collection geometry. **d,** Temporal full width at half maximum (FWHM) versus slit displacement at an MCP gain setting of 2. Gray open circles show individual valid measurements; red circles and error bars indicate the mean and sample standard deviation, respectively. The slit displacements are 0, 0.25, 0.40, 0.50, 0.60 and 0.70 mm.

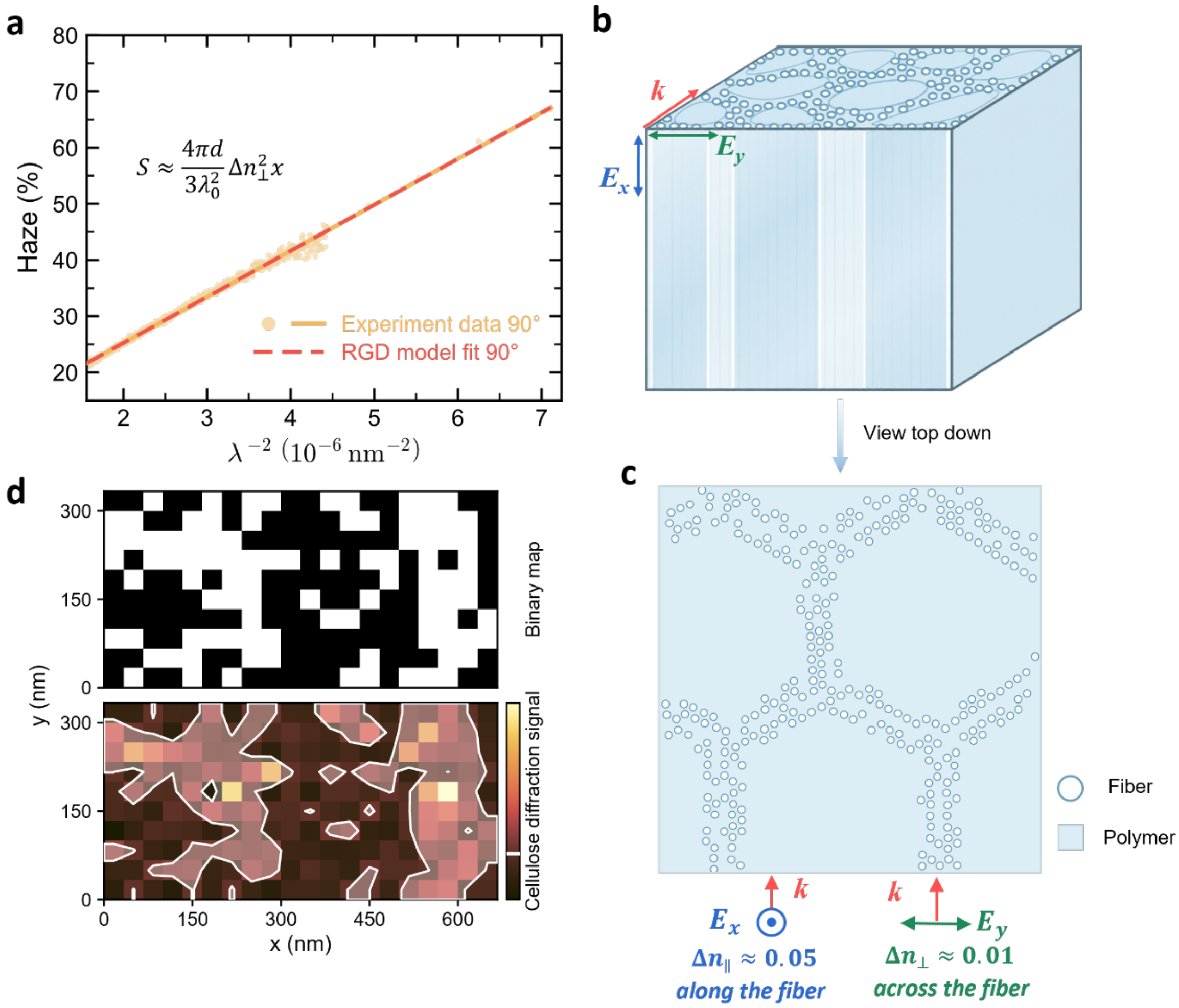

**Figure 4. Birefringence-related scattering in transparent wood. a,** Measured haze (scattering) at perpendicular polarization (90°) plotted against $\lambda^{-2}$ with the fit associated with the Rayleigh–Gans–Debye (RGD) parameterization ($R^2$ = 0.994). The fit includes a constant offset ($H_0$=0.08), while the displayed expression describes the inverse-square scattering contribution. **b,** Schematic of the polymer-infiltrated cellulose framework, defining the propagation direction $k$ and electric-field orientations $E_x$ and $E_y$. **c,** Top-view schematic illustrating cellulose–polymer refractive-index mismatches for polarization parallel ($\Delta n_\parallel \approx 0.05$) and perpendicular ($\Delta n_\perp \approx 0.01$) to the fibers. **d,** Continuous (bottom) and binary (top) cellulose diffraction maps derived from scanning electron diffraction, showing spatially heterogeneous diffraction signals within the cell wall. Continuous-map contours indicate a signal threshold. In the binary map, white and black indicate positions with and without detectable cellulose diffraction, respectively.

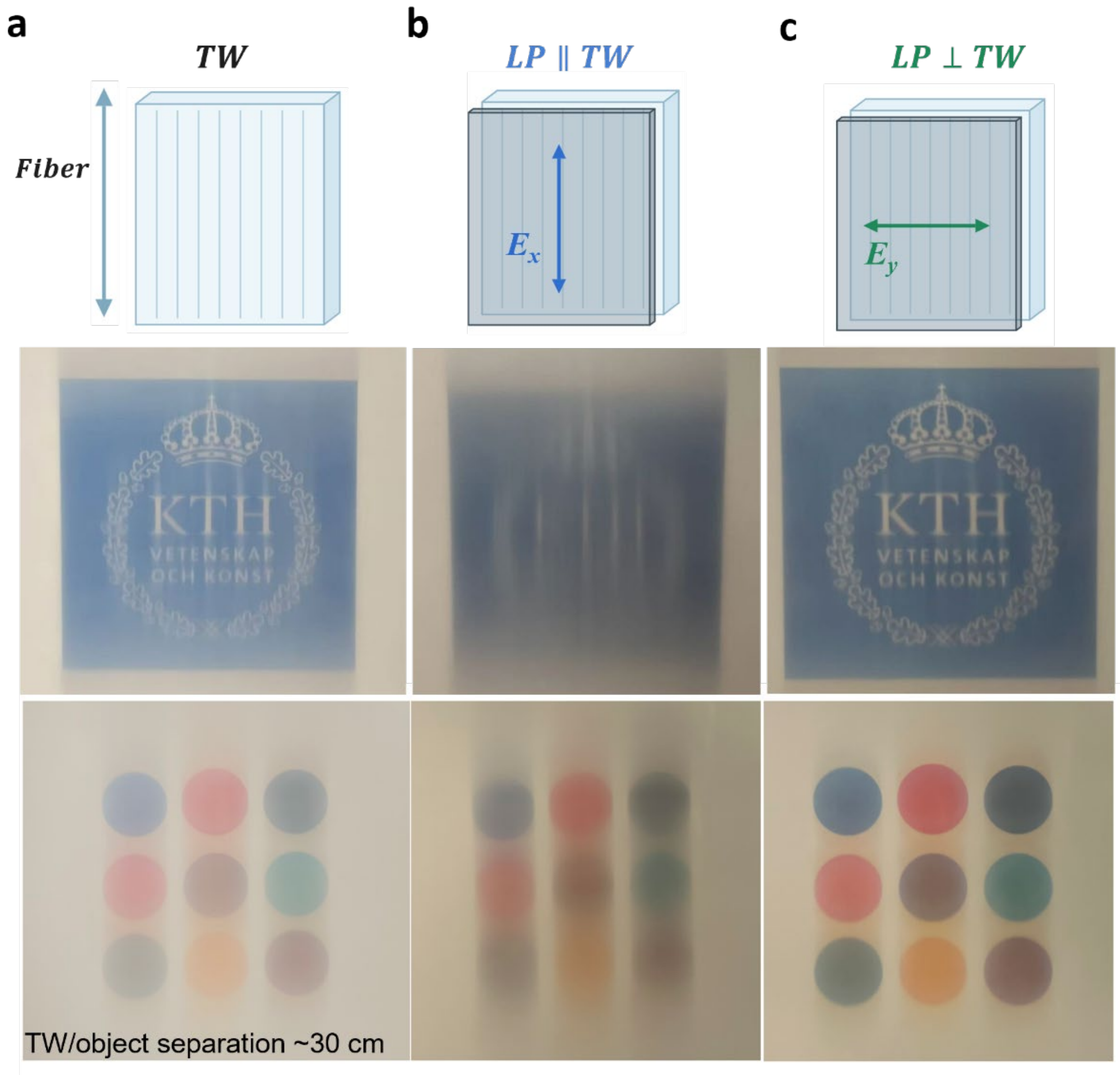

**Figure 5. Improved image clarity through polarization selection.** Photographs of a logo and color target viewed through (a) transparent wood alone and with the transmission axis of a linear polarizer (b) parallel or (c) perpendicular to the wood fibers. Perpendicular polarization improves image clarity at reduced brightness. The object-to-sample separation is ~30 cm.